\documentclass[fleqn,usenatbib]{mnras}

\usepackage[T1]{fontenc}
\usepackage{ae,aecompl}

\usepackage{graphicx}	% Including figure files
\usepackage{amsmath}	% Advanced maths commands
\usepackage{amssymb}	% Extra maths symbols
\usepackage{xspace}
\usepackage{hyperref}

\newcommand{\epic}{K2-99\xspace}
\newcommand{\ktwo}{\textit{K2}\xspace}
\newcommand{\kep}{\textit{Kepler}\xspace}
\newcommand{\rhk}{$\log R'_{\rm HK}$\xspace}
\newcommand{\feh}{\mbox{[Fe/H]}\xspace}
\newcommand{\teff}{\mbox{$T_{\rm *, eff}$}\xspace}
\newcommand{\logg}{\mbox{$\log g_*$}\xspace}
\newcommand{\vsini}{\mbox{$v \sin i_{*}$}\xspace}
\newcommand{\kms}{\mbox{km\,s$^{-1}$}\xspace}
\newcommand{\ms}{\mbox{m\,s$^{-1}$}\xspace}
\newcommand{\mplanet}{\mbox{$M_{\rm p}$}\xspace}
\newcommand{\rplanet}{\mbox{$R_{\rm p}$}\xspace}
\newcommand{\mjup}{\mbox{$M_{\rm Jup}$}\xspace}
\newcommand{\rjup}{\mbox{$R_{\rm Jup}$}\xspace}
\newcommand{\mstar}{\mbox{$M_{*}$}\xspace}
\newcommand{\rstar}{\mbox{$R_{*}$}\xspace}
\newcommand{\densstar}{\mbox{$\rho_*$}\xspace}
\newcommand{\msol}{\mbox{$M_\odot$}\xspace}
\newcommand{\rsol}{\mbox{$R_\odot$}\xspace}
\newcommand{\ecos}{\mbox{$e \cos \omega$}\xspace} 
\newcommand{\esin}{\mbox{$e \sin \omega$}\xspace}

\title[A transiting warm Jupiter around \epic]{\epic: a subgiant hosting a transiting warm Jupiter in an eccentric orbit and a long-period companion}

\author[A. M. S. Smith et al.]{
A. M. S. Smith$^{1}$\thanks{E-mail: Alexis.Smith@dlr.de},
D.~Gandolfi$^{2,3}$,
O.~Barrag\'{a}n$^{2}$,
B.~Bowler$^{4,5}$,
Sz.~Csizmadia$^{1}$,
\newauthor
M.~Endl$^{4}$,
M.~C.~V.~Fridlund$^{6,7}$,
S.~Grziwa$^{8}$,
E.~Guenther$^{9}$,
A.~P.~Hatzes$^{9}$,
\newauthor
G.~Nowak$^{10,11}$,
S.~Albrecht$^{12}$,
R.~Alonso$^{10,11}$,
J.~Cabrera$^{1}$,
W.~D.~Cochran$^{4}$,
\newauthor
H.~J.~Deeg$^{10,11}$,
F.~Cusano$^{13}$,
Ph.~Eigm\"{u}ller$^{1}$,
A.~Erikson$^{1}$,
D.~Hidalgo$^{10,11}$,
\newauthor
T.~Hirano$^{14}$,
M.~C.~Johnson$^{4,15}$,
J.~Korth$^{8}$,
A.~Mann$^{4}$,
N.~Narita$^{16,17,18}$,
\newauthor
D.~Nespral$^{10,11}$,
E.~Palle$^{10,11}$,
M. P\"{a}tzold$^{8}$,
J.~Prieto-Arranz$^{10,11}$,
H.~Rauer$^{1,19}$,
\newauthor
I.~Ribas$^{20}$,
B.~Tingley$^{12}$,
and V.~Wolthoff$^{3}$
\\
~\\
$^{1}$Institute of Planetary Research, German Aerospace Center, Rutherfordstrasse 2, 12489 Berlin, Germany\\
$^{2}$Dipartimento di Fisica, Universit\'{a} di Torino, via P. Giuria 1, 10125 Torino, Italy\\
$^{3}$Landessternwarte K\"{o}nigstuhl, Zentrum f\"{u}r Astronomie der Universit\"{a}t Heidelberg, K\"{o}nigstuhl 12, 69117 Heidelberg, Germany\\
$^{4}$Department of Astronomy and McDonald Observatory, University of Texas at Austin, 2515 Speedway, Stop C1400, Austin, TX 78712, USA\\
$^{5}$McDonald Prize Fellow\\
$^{6}$Leiden Observatory, University of Leiden, PO Box 9513, 2300 RA, Leiden, The Netherlands\\
$^{7}$Department of Earth and Space Sciences, Chalmers University of Technology, Onsala Space Observatory, 439 92 Onsala, Sweden\\
$^{8}$Rheinisches Institut f\"{u}r Umweltforschung, Abt. Planetenforschung, an der Universit\"{a}t zu K\"{o}ln, Aachener Strasse 209, 50931 K\"{o}ln, Germany\\
$^{9}$Th\"{u}ringer Landessternwarte Tautenburg, Sternwarte 5, 07778 Tautenburg, Germany\\
$^{10}$Instituto de Astrof\'{i}sica de Canarias, 38205 La Laguna, Tenerife, Spain\\
$^{11}$Departamento de Astrof\'{i}sica, Universidad de La Laguna, 38206 La Laguna, Tenerife, Spain\\
$^{12}$Stellar Astrophysics Centre, Department of Physics and Astronomy, Aarhus University, Ny Munkegade 120, DK-8000 Aarhus C, Denmark\\
$^{13}$INAF -- Osservatorio Astronomico di Bologna, Via Ranzani 1, 40127, Bologna, Italy\\
$^{14}$Department of Earth and Planetary Sciences, Tokyo Institute of Technology, 2-12-1 Ookayama, Meguro-ku, Tokyo 152-8551, Japan\\
$^{15}$Department of Astronomy, The Ohio State University, 140 West 18th Ave., Columbus, OH 43210, USA\\
$^{16}$Department of Astronomy, The University of Tokyo, 7-3-1 Hongo, Bunkyo-ku, Tokyo 113-0033, Japan\\
$^{17}$Astrobiology Center, National Institutes of Natural Sciences, 2-21-1 Osawa, Mitaka, Tokyo 181-8588, Japan\\
$^{18}$National Astronomical Observatory of Japan, 2-21-1 Osawa, Mitaka, Tokyo 181-8588, Japan\\
$^{19}$Center for Astronomy and Astrophysics, TU Berlin, Hardenbergstr. 36, 10623 Berlin, Germany\\
$^{20}$Institut de Ci\`encies de l'Espai (CSIC-IEEC), Carrer de Can Magrans, Campus UAB, 08193 Bellaterra, Spain\\
}

\date{Accepted XXX. Received YYY; in original form ZZZ}

\pubyear{2016}

\begin{document}
\label{firstpage}
\pagerange{\pageref{firstpage}--\pageref{lastpage}}
\maketitle

% Abstract of the paper
\begin{abstract}
We report the discovery from \ktwo of a transiting planet in an 18.25-d, eccentric ($0.19\pm 0.04$) orbit around \epic, an 11th magnitude subgiant in Virgo. We confirm the planetary nature of the companion with radial velocities, and determine that the star is a metal-rich ($\feh = 0.20\pm0.05$) subgiant, with mass $1.60^{+0.14}_{-0.10}$~\msol and radius $3.1\pm0.1$~\rsol. The planet has a mass of $0.97\pm0.09$~\mjup and a radius $1.29\pm0.05$~\rjup. A measured systemic radial acceleration of $-2.12\pm0.04~\mathrm{m~s^{-1} d^{-1}}$ offers compelling evidence for the existence of a third body in the system, perhaps a brown dwarf orbiting with a period of several hundred days.

\end{abstract}

% Select between one and six entries from the list of approved keywords.
% Don't make up new ones.

%%%%%%%%%%%%%%%%%%%%%%%%%%%%%%%%%%%%%%%%%%%%%%%%%%

%%%%%%%%%%%%%%%%% BODY OF PAPER %%%%%%%%%%%%%%%%%%

\section{Introduction}

Exoplanets that transit their host star are vital for our understanding of planetary systems, not least because their sizes and -- in combination with radial velocity (RV) measurements -- their absolute masses can be measured. Recent results from the CoRoT \citep{Moutou13} and \kep missions \citep{Borucki10} have both extended the parameter space of transiting planet discovery, particularly to longer orbital periods, and revolutionised our understanding of the planetary population of our Galaxy \citep[e.g.][]{Howard12}. A majority of the planets discovered by \kep, however, orbit stars too faint to enable RV measurements, and other observations, such as atmospheric characterisation, to be performed. Despite the great successes of \kep, most of the best-studied exoplanetary systems remain those discovered from the ground, by means of RV (in a few cases) or from surveys such as WASP \citep{Pollacco06} and HAT-net \citep{HATnet}.

The re-purposing of the \kep satellite to observe a number of fields along the ecliptic plane, for $\sim~80$~d each, the so-called \ktwo mission \citep{K2}, allows the gap between \kep and the ground-based surveys to be bridged. \ktwo  observes a large number of relatively bright ($V \lesssim 12$) stars, and has discovered a significant number of planets around such stars \citep[see][for a summary of the discoveries from \ktwo's first few fields]{Crossfield16}. \ktwo also allows the detection of smaller, and longer-period planets than are possible from the ground. The high-precision photometry achievable from space enables the discovery of small transit signals, and hence planets, as well as aiding the detection of long-period planets from just a few transits. The continuous nature of the observations eliminates the window functions associated with ground-based observations, and thus also helps to facilitate the discovery of relatively long-period systems.

\ktwo's great strengths then, are that it is capable of finding both bright planetary systems, and relatively long-period planets (at least by comparison with those discovered from the ground). The planetary system described in this paper, \epic, is a prime example of a system that is both bright ($V = 11.15$) and long-period ($P = 18.25$~d). To date only a handful of planets with periods longer than 10~d have been discovered by means of transits observed from the ground, and none with a period longer than that of \epic~b. \epic is one of a small number\footnote{about a dozen according to the {\it Exoplanet Data Explorer} (\citealt{exoplanets.org}; \url{http://www.exoplanets.org}).} of transiting planetary systems containing a planet on a long-period (> 10~d) orbit around a bright ($V < 12$) star.

Furthermore, \epic~b transits a star that is about to ascend the red-giant branch, and joins a small, but growing number of planets known to transit subgiant stars. In contrast to planets of solar-like stars, very little is known about planets of stars more massive than the Sun. This lack of knowledge is unfortunate, because theories of planet formation make very different predictions, whether such planets are rare or frequent \citep{Laughlin04, Ida&Lin05, K&K08, Alibert11, Hasegawa13, Kornet04, Boss06}. Thus, studies of the frequency of planets of stars more massive than the Sun are excellent tests of theories of planet formation. To date, most of the 156 known planet hosts more massive than 1.5~\msol are giant stars. According to the statistical analysis of \cite{Johnson10a, Johnson10b}, the frequency of massive planets increases with stellar mass. However, because all of the systems included in those analyses were detected by means of optical radial-velocity measurements, and because their orbit distribution is different from that of solar-like stars, and because there is also a lack of multiple planets, there are still some doubts as to whether the planets of giant stars are real \citep{Sato08, Lillo-Box16}. It is therefore necessary to confirm at least a few planets of stars more massive than the Sun by other methods. An important confirmation was the radial-velocity measurements in the near-IR recently carried out by \cite{Trifonov15}. The results for giant stars have furthermore been criticized in the sense that the masses of the giant stars could be wrong \citep{Lloyd13, Schlaufman13}.
 
The best confirmation would therefore be the detection of transiting planets of giant, or subgiant stars that are more massive than 1.5 \msol, or of main sequence A-stars for which it is also certain that they are more massive 1.5 \msol. Up to now 31 transiting planets of stars more massive than 1.5 \msol have been detected, but most of them are main sequence F-stars. A dedicated survey for transiting planets of A-stars with the CoRoT satellite turned up one planet around an F-star, and six A-star host candidates. The number of candidates corresponds to the expectations if the frequency of massive, close-in planets of A-stars were the same as that of G-stars \citep{Guenther16}. However, confirming these candidates is difficult given that the stars are faint, and rapidly rotating.
 
Only four planets are known to transit giant stars, and a further three transiting planets are known around subgiants (Section~\ref{sec:subgiant}). It is therefore of crucial importance to detect more transiting planets of giant, and subgiant stars with $\mstar > 1.5 \msol$, in order to find out whether planets, particularly short period ones, of such star are rare, or abundant. Here we present the discovery of a transiting planet around a subgiant of mass $1.60^{+0.14}_{-0.10}$~\msol.

\section{Observations}
\subsection{\ktwo Photometry}
\label{sec:phot}

\ktwo's Campaign 6 observations were centred on $\alpha\,=\,13^\mathrm{h}\,39^\mathrm{m}\,28^\mathrm{s}$ $\delta\,=\,-11\degr\,17\arcmin\,43\arcsec$ (2000.0) and ran from 2015 July 14 to 2015 September 30, i.e. for 78~d. A total of 28\,289 targets were observed in the standard 30-minute long-cadence mode, as well as 84 in short-cadence mode, and some custom targets.

\begin{figure}
	\includegraphics[width=\columnwidth]{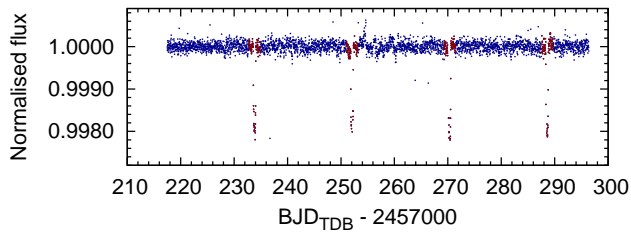}
    \caption{\ktwo light curve of \epic, processed by the {\tt K2SC} code of Aigrain et al. (2016), which removes both instrumental and stellar noise. Portions of the light curve selected for modelling are shown in red (Section~\ref{sec:anal}).}
    \label{fig:raw}
\end{figure}

EPIC~212803289 (later given the identifier K2-99) was identified as a candidate transiting planetary system from a search of \ktwo light curves extracted by \cite{vburg} performed using the {\tt EXOTRANS} pipeline along with the {\tt VARLET} filter \citep{exotrans,exotrans2}. Four transits, spaced every $\sim 18.25$~d, are clearly visible in the light curve of \epic (Fig. \ref{fig:raw}). On the basis of this detection (and the lack of odd -- even transit-depth variations, and the lack of a visible secondary eclipse) the system was selected for spectroscopic follow-up observations.

Independently, \epic was identified as a candidate by \cite{Pope16}. Using the {\tt K2SC} code of \cite{Aigrain16}, which relies on Gaussian processes to correct simultaneously the light curve for \ktwo pointing systematics and stellar variability, \cite{Pope16} identified a total of 152 candidate transiting systems from \ktwo Campaigns 5 and 6. The {\tt K2SC} light curve of \epic is shown in Fig.~\ref{fig:raw}, and is the light curve used in the rest of this work, as it appears to be marginally less noisy than that of \cite{vburg}\footnote{After submission of this paper, we became aware of {\tt EVEREST} \citep{EVEREST}, a \ktwo de-trending algorithm that produces a light curve with slightly less noise still.}.

%%%%%%%%%%%%%%%%% FULL  LIGHT  CURVE  FIGURE %%%%%%%%%%%%%%%%%%
%%%%%%%%%%%%%%%%% FULL  LIGHT  CURVE  FIGURE %%%%%%%%%%%%%%%%%%

\subsection{Spectroscopic observations}

In order to confirm the planetary nature of the transiting object and measure its mass, we performed intensive spectroscopic follow-up with the following spectrographs: FIES \citep{Frandsen1999,Telting2014}, mounted on the 2.56-m Nordic Optical Telescope (NOT) and HARPS-N \citep{HARPS-N}, mounted on the 3.58-m Telescopio Nazionale Galileo (TNG), both located at the Observatorio del Roque de los Muchachos, La Palma, Spain; HARPS \citep{Mayor-etal03}, on the ESO 3.6-m Telescope at La Silla, Chile; and the Robert G. Tull coud\'{e} spectrograph \citep{Tull1995} on the 2.7-m Harlan J. Smith Telescope at McDonald Observatory, Texas, USA. The resulting radial velocity measurements are listed in Table~\ref{tab:rv}.

\subsubsection{FIES}

We acquired fourteen FIES spectra between March and July 2016. The instrument was used in its \emph{high-res} mode, which provides a resolving power of R\,$\approx$\,67\,000 in the spectral range 364-736~nm. We followed the same observing strategy adopted by \citet{Buchhave2010} and \citet{Gandolfi2015}, i.e., we traced the RV drift of the instrument by acquiring long-exposed (T$_\mathrm{exp}$\,$\approx$\,35 sec) ThAr spectra immediately before and after each target observation. The exposure times were 2700--3600 sec, leading to a signal-to-noise ratio (S/N) of about 40-50 per per pixel at 550~nm. The data were reduced using standard \texttt{IRAF} and \texttt{IDL} routines. Radial velocity measurements were derived via S/N-weighted, multi-order cross-correlations with the RV standard star HD\,50692 -- observed with the same instrument set-up as \epic. 

\subsubsection{HARPS-N}

We acquired five HARPS-N high resolution spectra (R$\approx$115\,000) between April and May 2016, as part of the observing programmes A33TAC\_11, A33TAC\_15, and AOT33-11. We set the exposure time to 1800--2100 sec and monitored the sky background using the second fibre. The data were reduced using the dedicated HARPS-N data reduction software pipeline. The S/N of the extracted spectra is about 40-50 per pixel at 550~nm. Radial velocities (Table~\ref{tab:rv}) were extracted by cross-correlation with a G2 numerical mask.

\subsubsection{HARPS}

We also acquired eleven HARPS high resolution spectra (R$\approx$115\,000) between April and August 2016 under the ESO programme 097.C-0948. We set the exposure time to 1800--2100 sec, leading to a S/N of about 30-50 per pixel at 550~nm on the extracted spectra. We monitored the sky background using the second fibre and reduced the data with the HARPS data reduction software pipeline. Radial velocities (Table~\ref{tab:rv}) were extracted by cross-correlation with a G2 numerical mask. Three out of the eleven HARPS RVs are affected by technical problems and are not listed in (Table~\ref{tab:rv}). Nevertheless, the three HARPS spectra were used to derive the spectral parameters of \epic, as described in Section~\ref{sec:specanal}.

\subsubsection{Tull}

We obtained six precise RV measurements with the Tull Coud\'e spectrograph. The instrument covers the entire optical spectrum at a resolving power of R$\approx$60\,000. We used a molecular iodine (I$_2$) absorption cell for simultaneous wavelength calibration and point-spread function reconstruction. The differential RVs were calculated with our standard I$_2$-cell data modelling code \texttt{Austral} \citep{Endl2000}. For the stellar template we employed the co-added HARPS-N spectrum of \epic which has a sufficient high S/N of $\sim$100.

%%%%%%%%%%%%%%%%% RADIAL VELOCITY TABLE %%%%%%%%%%%%%%%%%%
\begin{table}
\caption{Radial velocity measurements, uncertainties and cross-correlation function bisector spans (BS) of \epic}
\begin{tabular}{crcrr}\hline
$\mathrm{BJD_{TDB}}$ & RV & $\sigma_{\mathrm{RV}}$ & BS & Instrument \\
 $-2450000$ & \kms & \kms  & \kms & \\
\hline
7479.624340  &  -2.697  &   0.013  & $+$0.013  & FIES\\
7492.520141  &  -2.581  &   0.008  & -0.007  &  HARPS-N\\
7493.757674  &  $+$0.502  &   0.019  & -- &  Tull\\
7494.804635  &  $+$0.467  &   0.016  & -- &  Tull\\
7502.643805  &  -2.493  &   0.007  & -0.031  &  HARPS-N\\
7503.531525  &  -2.601  &   0.014  & -0.003  &  FIES\\
7511.732010  &  -2.602  &   0.004  & -0.040  & HARPS\\
7512.508450  &  -2.616  &   0.005  & $+$0.007   & HARPS-N\\
7512.634721  &  -2.622  &   0.004  & -0.000   &  HARPS\\
7515.726524  &  -2.617  &   0.012  & $+$0.054   & HARPS\\
7516.569369  &  -2.598  &   0.006  & -0.053   & HARPS\\
7523.478018  &  -2.630  &   0.019  & -0.025   & FIES \\
7524.768623  &  $+$0.446  &   0.015  & -- &  Tull \\
7532.518735  &  -2.645  &   0.006  & -0.034  & HARPS-N\\
7539.461243  &  -2.558  &   0.005  & -0.019  & HARPS-N\\
7542.699191  &  $+$0.477  &   0.008  & -- & Tull\\
7543.736409  &  $+$0.433  &   0.011  & -- & Tull\\
7545.696704  &  $+$0.416  &   0.021  & -- & Tull\\
7559.601582  &  -2.620  &   0.005  & -0.023  & HARPS\\
7561.581344  &  -2.649  &   0.005  & -0.014  & HARPS\\
7565.410818  &  -2.806  &   0.016  & -0.030    & FIES\\
7566.413167  &  -2.798  &   0.014  & -0.000    & FIES\\
7567.416731  &  -2.849  &   0.014  & -0.008    & FIES\\
7568.417452  &  -2.834  &   0.018  & $+$0.030   & FIES\\
7570.405863  &  -2.819  &   0.016  & -0.020    & FIES\\
7572.408029  &  -2.809  &   0.016  & -0.018    & FIES\\
7575.409114  &  -2.740  &   0.018  & -0.014    & FIES\\
7576.403828  &  -2.726  &   0.015  & $+$0.003   & FIES\\
7577.404365  &  -2.757  &   0.020  & $+$0.007   & FIES\\
7578.405228  &  -2.761  &   0.016  & $+$0.004   & FIES\\
7579.402440  &  -2.807  &   0.022  & $+$0.021   & FIES\\
7589.495744  &  -2.779  &   0.006  & -0.050  &  HARPS\\
7610.468090  &  -2.770  &   0.005  & -0.025  &  HARPS\\
\hline
\\
\end{tabular}
\label{tab:rv}
\end{table}
%%%%%%%%%%%%%%%%% RADIAL VELOCITY TABLE %%%%%%%%%%%%%%%%%%

\subsection{Imaging}
\label{sec:ao}

In order to see if there exist close neighbours to \epic which could be diluting the transit signal, we performed adaptive-optics (AO) imaging of the target. We used the facility infrared imager NIRC2 at Keck Observatory using natural guide star adaptive optics \citep{Keck_AO} on 15 July 2016 UT. The narrow camera mode and $K_S$-band filter were chosen to finely sample the point spread function with a high Strehl ratio.  The resulting field of view was 10$\farcs$2$\times$10$\farcs$2. We acquired a set of ten short, unsaturated frames (10 coadds $\times$ 0.1 s each) and five deeper frames (1 coadd $\times$ 60 s each) behind the partly opaque 600 mas diameter coronagraph mask. Images were bias subtracted, flat fielded, and corrected for bad pixels and cosmic rays. \epic appears single down to the diffraction limit (FWHM = 46.3 $\pm$ 1.2 mas) and no point sources are evident in the deeper images.

\section{Analysis}
\label{sec:anal}

\subsection{Spectral analysis}
\label{sec:specanal}

We separately co-added the FIES, HARPS, and HARPS-N data to produce three master spectra that were used to derive the stellar parameters of \epic. We fitted the three master spectra to a grid of theoretical models from \citet{Castelli2004}, using spectral features that are sensitive to different photospheric parameters. We adopted the calibration equations for dwarf stars from \citet{Bruntt2010} and \citet{Doyle14} to determine the microturbulent, $v_ {\mathrm{micro}}$, and macroturbulent, $v_{\mathrm{macro}}$, velocities, respectively. The projected rotational velocity, \vsini, was measured by fitting the profile of several unblended metal lines. We also used the spectral analysis package \texttt{SME} (version 4.43) to perform an independent spectral analysis. \texttt{SME} calculates synthetic spectra of stars and fits them to observed high-resolution spectra \citep{Valenti1996,Valenti2005}. It solves for the model atmosphere parameters using a non-linear least squares algorithm. The two analyses provided consistent results well within the errors bars regardless of the used spectrum. The final adopted values are reported in Table~\ref{tab:stellar}.

%%%%%%%%%%%%%%%%% TRANSIT  LIGHT  CURVE  FIGURE %%%%%%%%%%%%%%%%%%
\begin{figure}
	\includegraphics[width=\columnwidth]{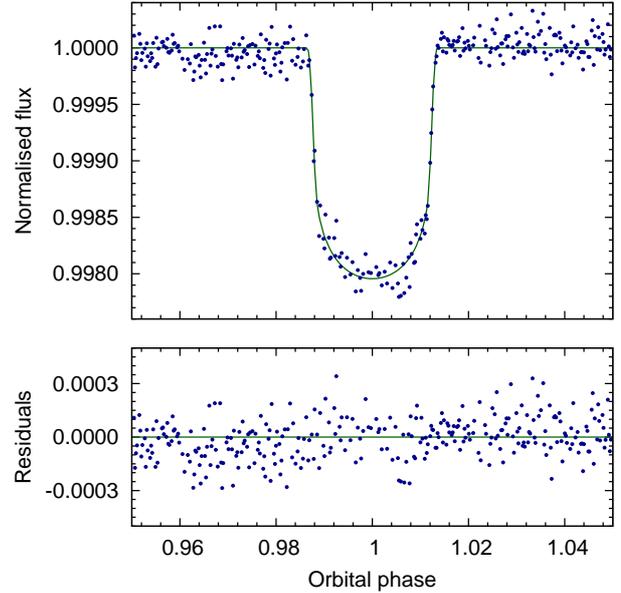}
    \caption{Phased light curve, overplotted with our best-fitting model.}
    \label{fig:transit}
\end{figure}
%%%%%%%%%%%%%%%%% TRANSIT  LIGHT  CURVE  FIGURE %%%%%%%%%%%%%%%%%%

%%%%%%%%%%%%%%%%% RADIAL VELOCITY FIGURE %%%%%%%%%%%%%%%%%%
\begin{figure}
	\includegraphics[width=\columnwidth]{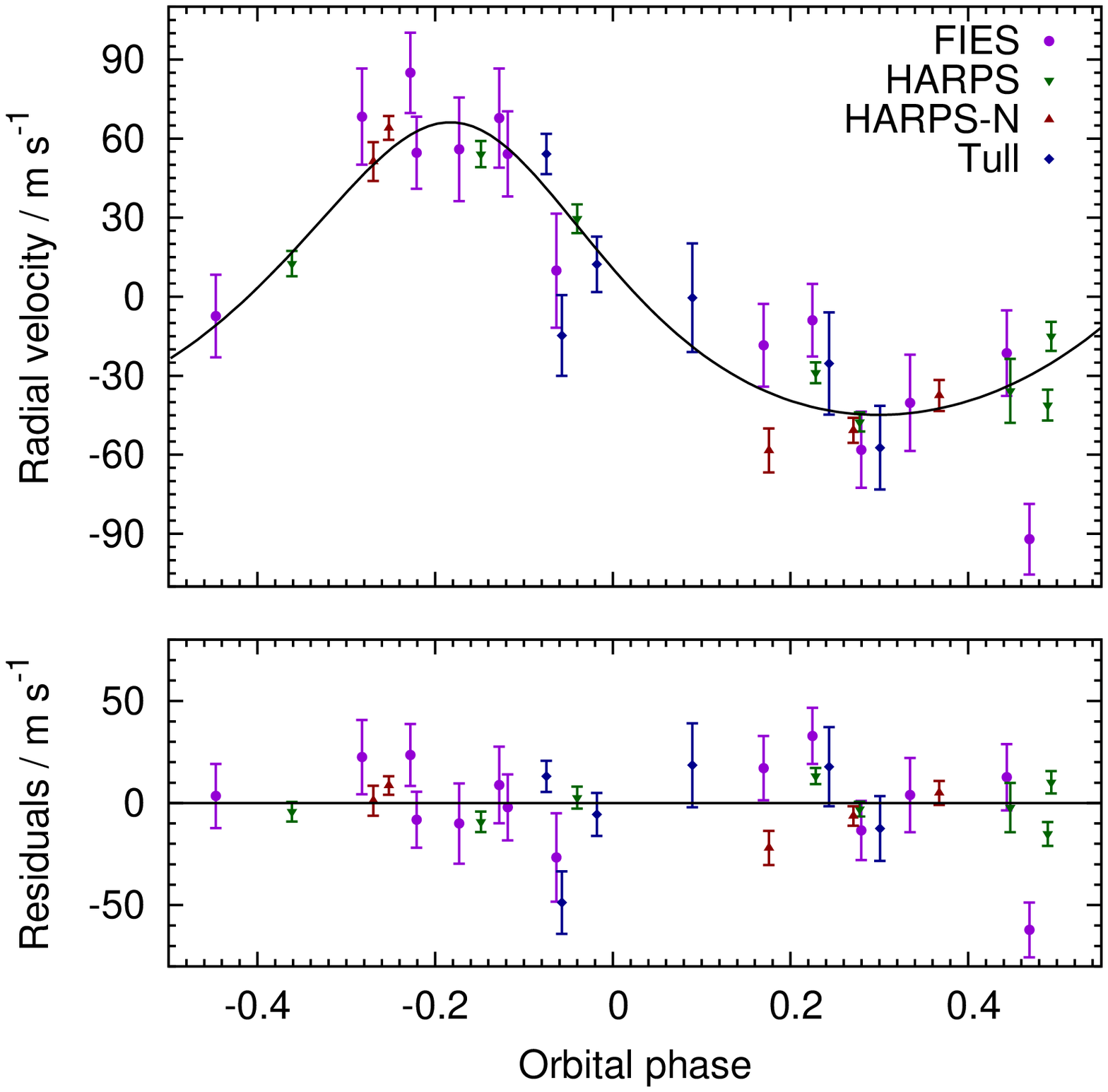}
	\includegraphics[width=\columnwidth]{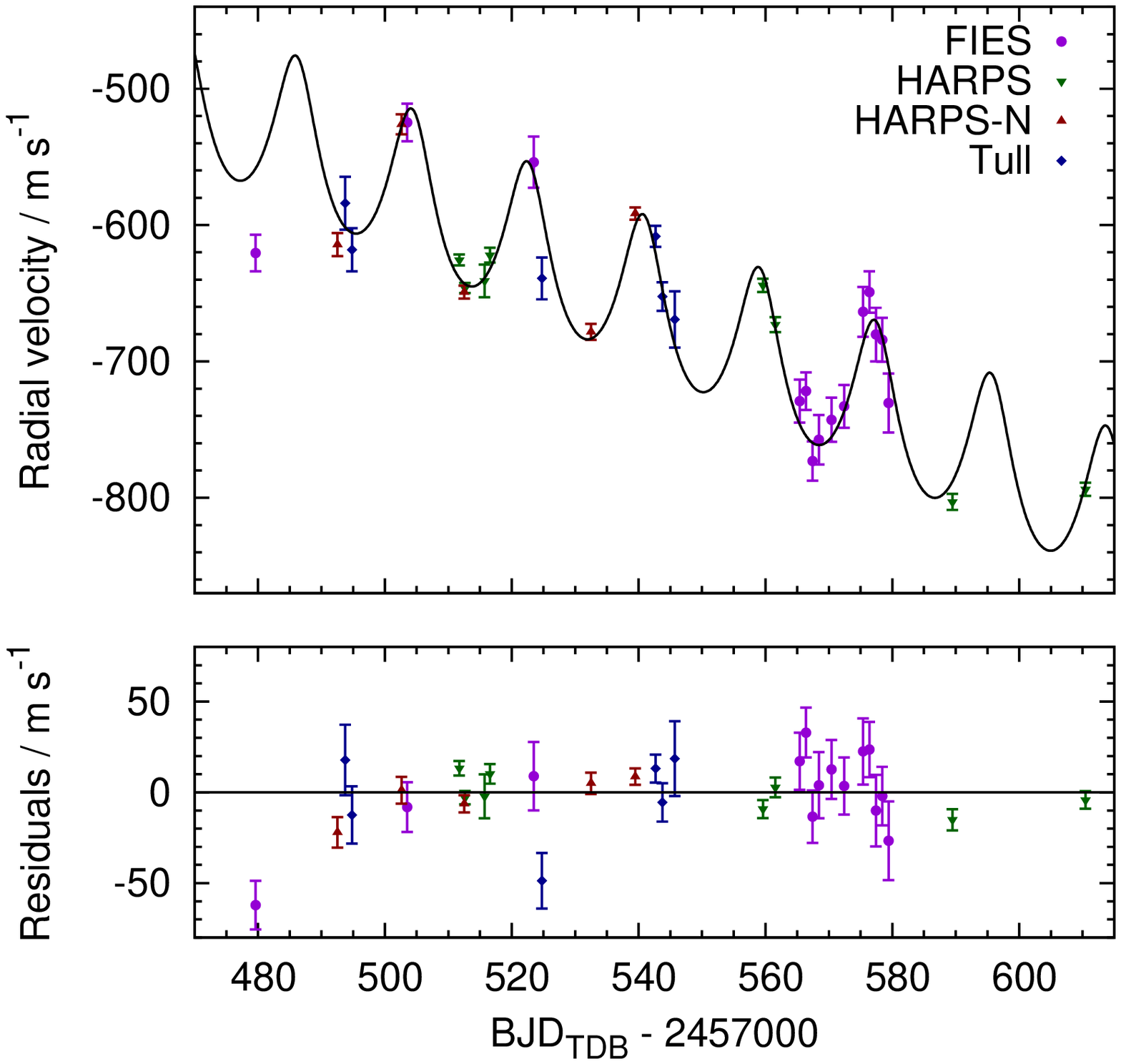}
    \caption{Radial velocities as a function of orbital phase (\textit{upper panels}) and time (\textit{lower panels}), with best-fitting models, and residuals to those models plotted below. Data points (with 1-$\sigma$ error bars) are from four different spectrographs, represented by different colours and symbol types.}
    \label{fig:rv}
\end{figure}
%%%%%%%%%%%%%%%%% RADIAL VELOCITY FIGURE %%%%%%%%%%%%%%%%%%

%%%%%%%%%%%%%%%%% RADIAL VELOCITY BISECTORS %%%%%%%%%%%%%%%%%%
\begin{figure}
    \includegraphics[width=\columnwidth]{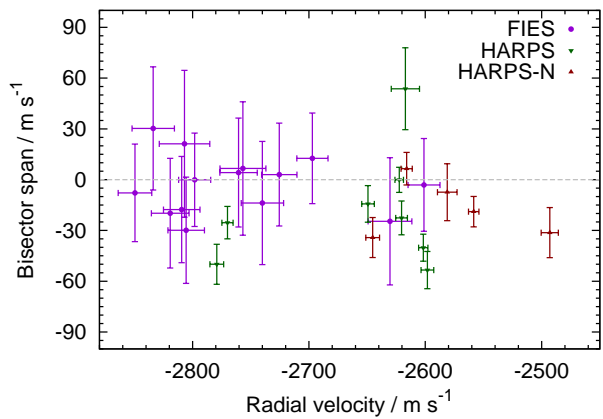}
    \caption{Radial velocity bisector span vs. relative radial velocity for data from the FIES, HARPS, and HARPS-N instruments. The uncertainties in the bisector spans are taken to be twice the uncertainty in the radial velocities.}
    \label{fig:bs}
\end{figure}
%%%%%%%%%%%%%%%%% RADIAL VELOCITY BISECTORS %%%%%%%%%%%%%%%%%%

%_________________Stellar parameters table________________
\begin{table}
\caption{Stellar parameters from our spectral analysis, and catalogue magnitudes for \epic}
\begin{tabular}{ll} \hline
Parameter  & Value \\ \hline
RA (J2000.0) & 13h55m05.7s  \\
Dec (J2000.0) & $-05^{\circ}~26\arcmin~32.88\arcsec $\\
\teff      &   $5990\pm40$ K \\
\logg (cgs)     &  $3.47\pm0.06$ \\
$v_ {\mathrm{micro}}$ & $1.2\pm0.1$ \kms \\
$v_ {\mathrm{macro}}$ & $5.8\pm0.6$ \kms \\
\vsini     &   $9.3\pm0.5$  \kms   \\
Spectral type   & G0 IV     \\
{[Fe/H]} & $0.20\pm0.05$\\
{[Ca/H]} & $0.30\pm0.05$\\
{[Mg/H]} & $0.25\pm0.05$\\
\hline
Magnitudes&(from EPIC$^\dagger$)\\
B (Tycho)&11.750 $\pm$ 0.113\\
g&11.332 $\pm$ 0.060\\
V (Tycho)&11.149 $\pm$ 0.099\\
r&10.957 $\pm$ 0.030\\
Kep&11.014\\
i&10.878 $\pm$ 0.040\\
J (2MASS)&10.024 $\pm$ 0.022\\
H (2MASS)&9.755 $\pm$ 0.021\\
K (2MASS)&9.720 $\pm$ 0.021\\
WISE 3.4 \micron&9.685 $\pm$ 0.021\\
WISE 4.6 \micron&9.714 $\pm$ 0.020\\
WISE 12 \micron&9.721 $\pm$ 0.047\\
WISE 22 \micron&8.850\\
\hline
\multicolumn{2}{l}{Additional identifiers for \epic:}\\
\multicolumn{2}{l}{TYC 4974-871-1}\\
\multicolumn{2}{l}{2MASS J13550570-0526330}\\
\multicolumn{2}{l}{EPIC~212803289}\\
\hline
$\dagger$\ktwo's Ecliptic Plane Input Catalog\\
\end{tabular}
\label{tab:stellar}
\newline {\bf References:} Tycho: \cite{Tycho}. 2MASS: \cite{2MASS}. WISE: \cite{WISE}.
\end{table}
%_________________Stellar parameters table________________

\subsection{Joint analysis of photometry and radial velocities}
\label{sec:analysis}

The photometry and radial velocities were analysed simultaneously using the current version of the {\tt Transit Light Curve Modelling (TLCM)} code (\citealt{Szilard_BD}; Csizmadia et al. in prep.). In brief, {\tt TLCM} uses the \cite{M&A} model to fit the transit photometry, whilst simultaneously fitting a Keplerian orbit to the RVs. A genetic algorithm is used to optimise the fit, and then a simulated annealing chain uses the output of the genetic algorithm as a starting point, and estimates the uncertainties over a large number of steps (typically $\sim10^5$). The Keplerian RV model is superimposed with a linear trend of radial velocity with time (see Section~\ref{sec:3rd}); we also fitted for an offset in RV between FIES and each of the other spectrographs.

The light curve of \epic we model is that generated by the {\tt K2SC} code of \cite{Aigrain16} (see Section~\ref{sec:phot}). We used only a subset of the light curve for modelling, selecting just over 1.5 times the transit duration both before and after each transit, such that the modelled light curve consists of four blocks of data, each around 2.2~d in duration, centred on each transit mid-point (see Fig.~\ref{fig:raw}). This has the effect of reducing the number of photometric data points from 3516 to 372. Because the effective \ktwo exposure time is relatively long (1800~s), we subdivide each exposure during the modelling, using a five-point Simpson integration.

The free parameters during the fitting process were the orbital period ($P$); the epoch of mid-transit ($T_\mathrm{c}$); the orbital major semi-axis in units of the stellar radius ($a/R_*$); the ratio of the planetary and stellar radii ($\rplanet / \rstar$); the orbital inclination angle ($i_\mathrm{p}$); the limb-darkening parameters, $u_+ = u_a + u_b$, and $u_- = u_a - u_b$, where $u_a$ and $u_b$ are the coefficients in a quadratic limb-darkening model; \esin and \ecos, where $e$ is the orbital eccentricity, and $\omega$ is the argument of periastron; the systemic radial velocity ($\gamma$); the stellar orbital velocity semi-amplitude ($K$); as well as the aforementioned radial acceleration ($\dot{\gamma}$) and instrumental RV offsets ($\gamma_{\rm 2-1}$, $\gamma_{\rm 3-1}$, and $\gamma_{\rm 4-1}$). The resulting fits to the transit photometry and the RVs are shown in Figs.~\ref{fig:transit} and \ref{fig:rv}, respectively.

The stellar mass and radius were calculated by comparing the mean stellar density, the stellar effective temperature, and the stellar metallicity to theoretical isochrones. The stellar density was measured from the joint fitting of the transit light curve and the RVs (Table~\ref{tab:mcmc}), and the stellar temperature and metallicity values are those derived in our spectral analysis (Section~\ref{sec:specanal}). We use the single star evolution ({\tt SSE}) isochrones of \cite{SSE}. 

Calculating the planetary radius is then trivial, since $\rplanet / \rstar$ is known. The planet mass, \mplanet, is calculated according to 
\begin{equation}
\mplanet \sin{i} = K \left(\frac{P}{2\pi G}\right)^{\frac{1}{3}} \mstar^{\frac{2}{3}}\sqrt{1 - e^2},
\end{equation}
given that $\mstar \gg \mplanet$.

The stellar mass and radius calculated from isochrones can be used to calculate the logarithm of the stellar surface gravity, $\logg = 3.67\pm0.04$. This value is in reasonably good agreement with that computed from our spectral analysis (Section~\ref{sec:specanal}, Table~\ref{tab:stellar}). The stellar age was determined to be $2.4^{+0.2}_{-0.6}$~Gyr.

%%%%%%%%%%%%%%%%% PARAMETERS TABLE %%%%%%%%%%%%%%%%%%
\begin{table*} 
\caption{System parameters from {\tt TLCM} modelling} 
\label{tab:mcmc}
\begin{tabular}{lccc}
\hline
\hline
Parameter & Symbol & Unit & Value\\
\hline 
\textit{{\tt TLCM} fitted parameters:} &\\
&\\
Orbital period	    	    	    	    & 	$P$ & d & $18.249\pm0.001$\\
Epoch of mid-transit	    	    	    & 	$T_{\rm c}$ &$\mathrm{BJD_{TDB}}$ & $2457233.823\pm0.003$\\
Scaled orbital major semi-axis     &   $a/R_{\rm *}$ &...& {}$11.1\pm0.1${}\\
Ratio of planetary to stellar radii  &  \rplanet / \rstar &...& $0.0422\pm0.0006$ \\
Orbital inclination angle   	    	    & 	$i_\mathrm{p}$ &$^\circ$  & $ 87.7\pm0.3$\\
...						    & \esin &...& $ 0.03\pm0.03$ \\
...						    & \ecos &...& $ 0.19\pm0.04$ \\
Limb-darkening parameters          & $u_+$ &...&$ 0.6\pm0.1$ \\
						    & $u_-$ &...&$ 0.08\pm0.20$ \\
Stellar orbital velocity semi-amplitude & 	$K$ &m s$^{-1}$ & $56 \pm4 $\\
Systemic radial velocity     	    	 &     	$\gamma$ &km s$^{-1}$ & $-2.08 \pm 0.01$\\
Systemic radial acceleration          &$\dot{\gamma}$&m s$^{-1}$ d$^{-1}$& $-2.12 \pm 0.04$\\
Velocity offset between FIES and HARPS & $\gamma_{\rm 2-1}$ & m s$^{-1}$ & $  100\pm 8$\\
Velocity offset between FIES and HARPS-N & $\gamma_{\rm 3-1}$ & m s$^{-1}$ & $  110\pm 7$\\
Velocity offset between FIES and Tull & $\gamma_{\rm 4-1}$ & m s$^{-1}$ & $  316\pm 12$\\
&\\
\textit{Derived parameters:}\\
&\\
Orbital eccentricity	    	    	    & 	$e$ &...& $0.19\pm0.04$ \\
Argument of periastron$^\dagger$                  &  $\omega$ & $^\circ$ & $8\pm8$\\
\noalign{\smallskip}  
Stellar mass	    	    	    	    & 	\mstar  & \msol  & $ 1.60^{+0.14}_{-0.10}  $\\
\noalign{\smallskip}  
Stellar radius	    	    	    	    & 	\rstar & \rsol & $ 3.1\pm0.1  $\\
log (stellar surface gravity)     	    & 	$\log g_{*}$ & (cgs) & $ 3.67\pm0.04  $\\
Stellar density     	    	    	    & 	 \densstar &kg m$^{-3}$ & $ 78\pm3  $\\
Planet mass 	    	    	    	    & 	\mplanet &\mjup & $ 0.97\pm0.09  $\\
Planet radius	    	    	    	    & 	\rplanet &\rjup & $1.29 \pm0.05  $\\
log (planet surface gravity)     	    & 	$\log g_{\rm p}$ & (cgs) & $ 3.2\pm0.1  $\\
Orbital major semi-axis     	    	    & 	$a$ &AU  & $ 0.159\pm0.006  $\\
Transit impact parameter               & $b$ &...& $0.41\pm0.05$\\
Transit duration                              & $T_{\rm 14}$ &d&$0.50\pm0.01$\\
\noalign{\smallskip}  
Stellar age				    & $\tau$ & Gyr & $2.4^{+0.2}_{-0.6}$\\
Distance (see Section~\ref{sec:dist})	     &  $d$ & pc&  $606\pm32$ \\
\hline
\end{tabular} \\ 
\end{table*} 
%%%%%%%%%%%%%%%%% PARAMETERS TABLE %%%%%%%%%%%%%%%%%%

\subsection{Orbital eccentricity}

In addition to fitting for \esin and \ecos, when we found $e = 0.19 \pm 0.04$, we also tried fitting a circular orbit by fixing \esin = \ecos = 0. Using the F-test approach of \cite{lucy_sweeney}, we find that there is a only a very small ($\approx 2\times10^{-4}$) probability that the apparent orbital eccentricity could have been observed if the underlying orbit were actually circular. We therefore conclude that the eccentricity we detect in the orbit of \epic~b is significant.

\subsection{Radial velocity bisectors \& stellar activity}

For the radial velocity measurements obtained with FIES, HARPS, and HARPS-N, we were able to measure the bisector spans. A correlation between the bisector spans and RV is indicative of a blended eclipsing binary system, or of RV variation as a result of stellar activity \citep{Queloz01}. As expected for a true planetary system, however, we see no significant correlation between the bisector spans and RV (Fig.~\ref{fig:bs}).

Furthermore, we observed no correlation between the RVs and either the corresponding FWHM values, or the \rhk activity index values (HARPS data only). The mean \rhk is $-5.2$  which, along with an apparent lack of photometric variability, is strongly suggestive of a relatively inactive star. 

\subsection{Reddening and stellar distance}
\label{sec:dist}

We followed the method outlined in \citet{Gandolfi08} to estimate the interstellar reddening ($A_\mathrm{v}$) and distance $d$ to the star. Briefly, $A_\mathrm{v}$ was derived by simultaneously fitting the observed colors (Table~\ref{tab:stellar}) with synthetic magnitudes computed from the \texttt{NEXTGEN} \citep{Hauschildt99} model spectrum with the same spectroscopic parameter as \epic. We assumed a normal value for the total-to-selective extinction ($R_\mathrm{v}=A_\mathrm{v}/E(B-V)=3.1$) and adopted the interstellar extinction law of \citet{Cardelli89}. The spectroscopic distance to the star was estimated using the de-reddened observed magnitudes and the \texttt{NEXTGEN} synthetic absolute magnitudes for a star with the same spectroscopic parameters and radius as \epic. We found that $A_\mathrm{v}=0.05\pm0.05$~mag and $d=606\pm32$~pc.

\subsection{Evidence for a third body}
\label{sec:3rd}

\subsubsection{Observed radial acceleration}
\label{sec:trend}

%%%%%%%%%%%%%%%%% TWO PLANET RV FIT  FIGURE %%%%%%%%%%%%%%%%%%
\begin{figure}
	\includegraphics[width=\columnwidth]{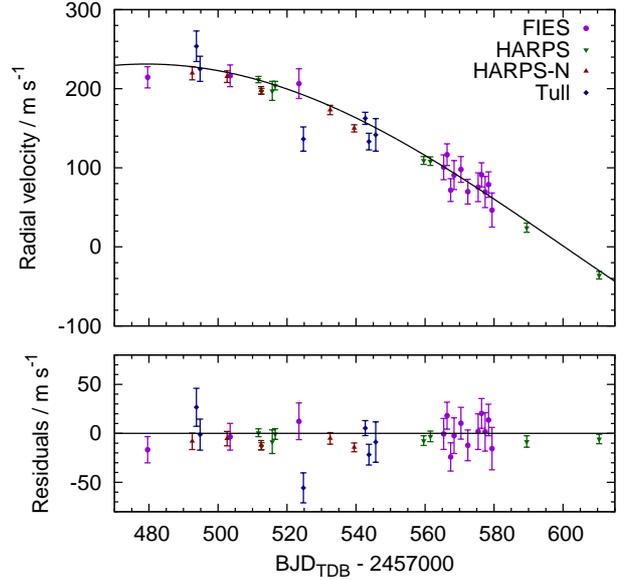}
    \caption{Residuals to the radial velocity fit for \epic~b, overplotted with a second fitted Keplerian orbit ($e=0$, $P_\mathrm{c} = 485\pm310$~d, $K_\mathrm{c}=230\pm150~\ms$). See Section~\ref{sec:trend} for details. The residuals to the double-Keplerian fit are plotted in the lower panel.}
    \label{fig:2p}
\end{figure}
%%%%%%%%%%%%%%%%% TWO PLANET RV FIT  FIGURE %%%%%%%%%%%%%%%%%%

We tried fitting the radial velocities both with and without the inclusion of a linear trend in time, finding that such a trend is heavily favoured by the data. Using the approach of \cite{Bowler16} (which follows \citealt{Torres99} and \citealt{Liu02}), we can place the following constraint on the properties of the third body, denoted `$c$':
\begin{equation}
\frac{M_\mathrm{c}}{a_\mathrm{c}^2} > 0.0145 \left \lvert \frac{\dot{\gamma}}{\mathrm{m}~\mathrm{s}^{-1}~\mathrm{yr}^{-1}} \right \rvert  = 11~ \mjup~\mathrm{au^{-2}}
\end{equation}

Furthermore, if we assume that the orbit of the third body is not significantly eccentric, we can infer that the period of the orbit must be at least twice the baseline of our RV data ($P_\mathrm{c} > 236$~d). This leads to the constraints that $a_\mathrm{c} \gtrsim 1.4~\mathrm{au}$, and hence $M_\mathrm{c} > 22~\mjup$. The likeliest possibilities, then, are a brown dwarf orbiting within about 2.7~au; a $\sim$\msol object at $\sim 10$~au; or an object orbiting on a highly-eccentric orbit, such that we have just observed the portion of the orbit where the induced stellar RV is greatest.

Noting that the RV model described above does not fit the very first RV point well, we decided to fit the RVs using the {\tt RVLIN} code and associated uncertainty estimator \citep{rvlin, rvlin_bootstrap}. The parameters we obtained for a one-planet fit with a constant radial acceleration are in excellent agreement with those obtained using {\tt TLCM} (Section~\ref{sec:analysis}). We then used {\tt RVLIN} to fit a second planet to the RVs, instead of a radial acceleration term. Unsurprisingly, the fit to the second planet is poorly constrained, but if we assume a circular orbit for the second planet, we find $P_\mathrm{c} = 485\pm310$~d and the orbital velocity amplitude due to the third body, $K_c = 230\pm150$~\ms (Fig.~\ref{fig:2p}). These values are used to calculate the minimum mass, $M_\mathrm{c} \sin i_\mathrm{c} = 14\pm9~\mjup$ and orbital major semi-axis, $a_\mathrm{c} = 1.4\pm1.0$~au. The two-planet fit results in a significantly lower $\chi^2$ than the linear acceleration model, and also a lower BIC (accounting for the increased number of free parameters in the two-planet model). We note, however, that favouring the two-planet model over the constant radial acceleration model relies heavily (but not entirely) on a single data point, our first RV measurement, and therefore caution against over-interpretation of the two-planet fit.

\subsubsection{AO imaging}
\label{sec:ao_res}

%%%%%%%%%%%%%% ADAPTIVE OPTICS SENSITIVITY MAP FIGURE %%%%%%%%%%%%%%%
\begin{figure}
	\includegraphics[width=\columnwidth]{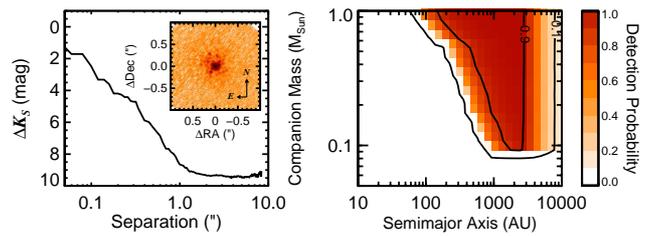}
    \caption{Detection limits for a luminous third body from NIRC2 imaging. The left panel shows the contrast curve generated by combining our shallow unsaturated images (\textit{inset}) with the deeper coronagraphic images.  The right panel shows the associated sensitivity to luminous third bodies (see Section~\ref{sec:ao_res} for details).
}
    \label{fig:ao}
\end{figure}
%%%%%%%%%%%%%% ADAPTIVE OPTICS SENSITIVITY MAP FIGURE %%%%%%%%%%%%%%%

Contrast curves and sensitivity maps from our NIRC2 observations are generated in the same manner as described in \cite{PALMS_4}.  Unsaturated and coronagraphic images are first corrected for optical distortions using the distortion solution from \cite{Service16}; then the images were registered, de-rotated to a common position angle to account for slight rotation in pupil-tracking mode, median-combined, and north-aligned using the \cite{Service16} north correction.  7-$\sigma$ contrast curves are generated using the rms in annuli centered on \epic together with the $K_S$-band coronagraph throughput measurement from \citep{PALMS_4}.  Finally, sensitivity maps are derived by generating artificial companions on random circular orbits and comparing their apparent magnitudes at the distance and age of \epic based on the evolutionary models of \cite{Baraffe15} to our contrast curve.  We also account for the fractional field of view coverage from the finite NIRC2 square detector. The resulting detection limits and sensitivity map are shown in Fig.~\ref{fig:ao}.

Unfortunately, given the large distance to \epic ($606\pm 32$~pc), the only limits we can place on the presence of a third body from AO imaging are at rather large distances from the star ($\gtrsim 100$~au). Both the observed radial acceleration and the second fitted Keplerian orbit (Section~\ref{sec:trend}), however, suggest that the third body is closer to \epic than that. The radial acceleration alone suggests that a star orbiting at 100~au would need to be very massive ($\approx 100 $~\msol) to fit the observations. In other words, the AO imaging does not help us to distinguish between the various possible scenarios identified in Section~\ref{sec:trend}.

\subsection{Other effects visible in the light curve}

We conducted searchs for, and placed upper limits on various photometric effects besides the planetary transits:

\subsubsection{Transit timing variations (TTV)}
We fitted for the epoch of each transit individually, using {\tt TLCM} to fit only the part of the light curve corresponding to a single transit, and keeping all parameters fixed to their best-fitting values (Table~\ref{tab:mcmc}), except for $T_{\rm c}$ which was allowed to vary. The individual times of mid-transit are reported in Table~\ref{tab:ttv}, and we see no evidence for any TTV. This non-detection is consistent with a maximum predicted TTV of 55~s, calculated from Equation 32 of \cite{Borkovits11}, using the third body parameters from Section~\ref{sec:trend} and further assuming that the mutual inclination angle between the two orbital planes is zero. We also see no compelling evidence of transit depth or profile variations, such as those caused by star-spot crossing events.

\subsubsection{Planetary occultation and phase variation}

We tried fitting for an occultation (secondary eclipse), and determine a best-fitting depth of $22\pm192$~ppm, we therefore place an upper limit (95 per cent confidence) on the occultation depth of 405~ppm. Similarly, we see no evidence for any orbital phase variation.

\subsubsection{Stellar rotational modulation}

We searched for evidence of stellar rotational modulation using {\tt Period04} \citep{period04}. We used the light curve of \cite{vburg}, since stellar variability is removed by the {\tt K2SC} pipeline. We found no evidence of such variability above an amplitude of $2 \times 10^{-5}$ (95 per cent confidence limit).

\subsubsection{Additional transits}

We searched the light curve for additional transits using the {\tt DST} code of \cite{DST}, but found no significant peaks in the periodogram which could indicate the existence of an additional transiting body.

%%%%%%%%%%%%%%%%% TRANSIT TIMINGS TABLE %%%%%%%%%%%%%%%%%%
\begin{table}
\caption{Fitted times of mid-transit for individual transits of \epic, their uncertainties and the deviation from the ephemeris presented in Table~\ref{tab:mcmc}}
\begin{tabular}{cccc}\hline
No. &$T_{\rm c} - 2\ 450\ 000$ & $\sigma_{T_{\rm c}}$ & O-C \\
&$\mathrm{BJD_{TDB}}$ & d & d \\
\hline
1  &     7233.8252   &    0.0015   &  +0.0017\\
2  &     7252.0708   &    0.0012   &  -0.0020\\
3  &     7270.3223   &    0.0010   &  +0.0001\\
4  &     7288.5722   &    0.0015   &  +0.0007\\
\hline
\\
\end{tabular}
\label{tab:ttv}
\end{table}
%%%%%%%%%%%%%%%%% TRANSIT TIMINGS TABLE %%%%%%%%%%%%%%%%%%

\section{Discussion and conclusions}

\subsection{\epic as a subgiant planet host star}
\label{sec:subgiant}

\epic joins a relatively short list of subgiants known to host transiting planets. The evolutionary tracks used to determine the stellar mass, radius and age (Section~\ref{sec:analysis}; \citealt{SSE}) suggest that the planet will be engulfed in around 150~Myr, as \epic expands further.

There have been several recent discoveries of transiting planets around subgiants, namely the short-period KELT-11b \citep[][$P=4.7$~d]{kelt11} and K2-39b \citep[][$P=4.6$~d]{k2-39}, which also shows evidence for a long-period companion. \epic is most reminiscent, however, of the Kepler-435 system (= KOI-680; \citealt{Almenara15}), which consists of an F9 subgiant ($\rstar = 3.2\pm0.3$) orbited by a giant planet in a slightly eccentric ($e=0.11\pm0.08$) 8.6-d orbit. Kepler-435 also exhibits a radial acceleration, most likely due to a planetary mass object in a $P > 790$~d orbit.

\subsection{\epic~b as a warm Jupiter}

\cite{Huang16} note that there appears to be a distinction between hot ($P<10$~d) and warm ($10<P<200$~d) Jupiters in that the latter are much more likely to have sub-Jovian companion planets. They find that around half of warm Jupiters (WJ) have smaller companions orbiting close to them, whereas this is true for only WASP-47 \citep{w47,w47_k2} amongst the hot Jupiters. We find no evidence for the existence of any sub-Jovian companions to \epic although we note that we are less sensitive to small planets (because of the large stellar radius) and long-period planets (because of \ktwo's limited observing baseline) than the \kep systems analysed by \cite{Huang16}. \epic also fits the correlation observed by \cite{Dawson13} that the orbits of WJs around metal-rich stars ($\feh \ge 0$, like \epic) have a range of eccentricities, whereas metal-poor stars host only planets on low-eccentricity orbits.

\subsection{Possible migration scenarios for \epic~b}

Using Equation (1) of \cite{Jackson08}, the current stellar parameters, and assuming $a$ to be constant, we calculate the circularisation time-scale, $\tau_e = \left(\frac{1}{e} \frac{de}{dt}\right)^{-1}$, for the orbit of \epic~b, in terms of the tidal dissipation parameters for the planet, $Q_{\rm P}$, and for the star, $Q_{\rm *}$,
\begin{equation}
\tau_e = \left(\frac{0.0104}{\left (\frac{Q_{\rm P}}{10^{5.5}}\right)} + \frac{0.0015}{\left (\frac{Q_{\rm *}}{10^{6.5}}\right)}\right)^{-1}~\mathrm{Gyr,}
\end{equation}
Adopting $Q_{\rm P} = 10^{5.5}$ and $Q_{\rm *} = 10^{6.5}$ (the best-fitting values from the study of \citealt{Jackson08}), we find $\tau_e = 84$~Gyr. Even in the case that $Q_{\rm P} = 35\,000$ (the value for Jupiter; \citealt{Lainey09}), and the extreme case that $Q_{\rm *} = 10^{5}$, the circularisation time-scale is still as long as 7.1~Gyr. These ages are much larger than the age of the system ($2.4^{+0.2}_{-0.6}$~Gyr), suggesting that the orbital eccentricity we observe is unlikely to have been significantly reduced by tidal interactions between the planet and star.

\cite{Dong14} note that a greater fraction of eccentric warm Jupiter systems contain a third body capable of having caused the inward migration of the WJ via a high-eccentricity mechanism. Although the orbital eccentricity of \epic~b is less than the threshold of 0.4 used by \cite{Dong14} to demarcate high-eccentricity systems, the system does contain such a potential perturber. WJs with observed eccentricities less than 0.4, however, may merely be at a low-$e$ stage in the cycle, and their orbits may become highly eccentric over a secular timescale. If \epic~b has undergone migration via a high-eccentricity route, such as Kozai migration, then one would expect the axis of its orbit to be significantly inclined with respect to the stellar spin axis (for it to have a large obliquity angle). We predict that the Rossiter-McLaughlin (R-M) effect for this system will have an amplitude of $\sim 11~\ms$. Given that, and the \vsini of $9.3\pm0.5$ ~\kms, it should be possible to detect the R-M effect, and measure the sky-projected obliquity for this system. To date, only seven WJs ($P>10$~d, $\rplanet > 0.6 \rjup$) have measured sky-projected obliquities, four of which are aligned and three of which show significant misalignment\footnote{Statistics are from Ren\'{e} Heller's Holt-Rossiter-McLaughlin Encyclopaedia, accessed on 2016 August 17 (\url{http://www.astro.physik.uni-goettingen.de/\textasciitilde rheller}).}. Further, \cite{Petrovich&Tremaine} predict that the companions to WJs should have high mutual inclination angles than those of hot Jupiters, typically 60\degr -- 80\degr.

\section*{Acknowledgements}
We wish to thank Trent Dupuy (University of Texas) for performing the AO observations. This paper includes data collected by the Kepler mission. Funding for the Kepler mission is provided by the NASA Science Mission directorate. Some of the data presented in this paper were obtained from the Mikulski Archive for Space Telescopes (MAST). STScI is operated by the Association of Universities for Research in Astronomy, Inc., under NASA contract NAS5-26555. Support for MAST for non-HST data is provided by the NASA Office of Space Science via grant NNX09AF08G and by other grants and contracts. Sz. Cs. acknowledges the Hungarian OTKA Grant K113117. N.~N. acknowledges a support by Grant-in-Aid for Scientific Research (A) (JSPS KAKENHI Grant Number 25247026). We are very grateful to the NOT, ESO, TNG and McDonald staff members for their unique support during the observations. We are also very thankful to Jorge Melendez, Martin K\"urster, Nuno Santos, Xavier Bonfils, and Tsevi Mazeh who kindly agreed to exchange HARPS and FIES time with us. Based on observations obtained \emph{a}) with the Nordic Optical Telescope (NOT), operated on the island of La Palma jointly by Denmark, Finland, Iceland, Norway, and Sweden, in the Spanish Observatorio del Roque de los Muchachos (ORM) of the Instituto de Astrof\'isica de Canarias (IAC); \emph{b}) with the Italian Telescopio Nazionale Galileo (TNG) also operated at the ORM (IAC) on the island of La Palma by the INAF - Fundaci\'on Galileo Galilei. Based on observations made with ESO Telescopes at the La Silla Observatory under programme ID 097.C-0948. This paper includes data taken at McDonald Observatory of the University of Texas at Austin. The research leading to these results has received funding from the European Union Seventh Framework Programme (FP7/2013-2016) under grant agreement No. 312430 (OPTICON) and from the NASA K2 Guest Observer Cycle 1 program under grant NNX15AV58G to The University of Texas at Austin. This research has made use of NASA's Astrophysics Data System, the SIMBAD database, operated at CDS, Strasbourg, France, the Exoplanet Orbit Database and the Exoplanet Data Explorer at exoplanets.org, the Exoplanets Encyclopaedia at exoplanet.eu, and René Heller's Holt-Rossiter-McLaughlin Encyclopaedia (www.astro.physik.uni-goettingen.de/\textasciitilde rheller). We thank the anonymous referee for comments which helped to improve this manuscript.

%%%%%%%%%%%%%%%%%%%%%%%%%%%%%%%%%%%%%%%%%%%%%%%%%%

%%%%%%%%%%%%%%%%%%%% REFERENCES %%%%%%%%%%%%%%%%%%

% The best way to enter references is to use BibTeX:

\bibliographystyle{mnras}
\bibliography{../refs2} % if your bibtex file is called example.bib

% Alternatively you could enter them by hand, like this:
% This method is tedious and prone to error if you have lots of references
%\begin{thebibliography}{99}
%\bibitem[\protect\citeauthoryear{Author}{2012}]{Author2012}
%Author A.~N., 2013, Journal of Improbable Astronomy, 1, 1
%\bibitem[\protect\citeauthoryear{Others}{2013}]{Others2013}
%Others S., 2012, Journal of Interesting Stuff, 17, 198
%\end{thebibliography}

%%%%%%%%%%%%%%%%%%%%%%%%%%%%%%%%%%%%%%%%%%%%%%%%%%

%%%%%%%%%%%%%%%%% APPENDICES %%%%%%%%%%%%%%%%%%%%%

%\appendix
%
%\section{Some extra material}
%
%If you want to present additional material which would interrupt the flow of the main paper,
%it can be placed in an Appendix which appears after the list of references.

%%%%%%%%%%%%%%%%%%%%%%%%%%%%%%%%%%%%%%%%%%%%%%%%%%

% Don't change these lines
\bsp	% typesetting comment
\label{lastpage}
\end{document}